\documentclass{ptapap}

\author{Slavek Rucinski}[DAA]
\author{Andrzej Pigulski}[Wroclaw]
\author{Adam Popowicz}[Gliwice]
\author{Rainer Kuschnig}[Graz,Vienna]
\author{Kre\v{s}imir Pavlovski}[Zagreb]
\author{the BRITE Team}

\affil[DAA]{Department of Astronomy and Astrophysics,
    University of Toronto, 50.\ St.George St., Toronto, Ontario, Canada M5S~3H4}
\affil[Wroclaw]{Instytut Astronomiczny, Uniwersytet Wroc{\l}awski, 
    Kopernika~11, 51-622 Wroc{\l}aw, Poland}
\affil[Gliwice]{Instytut Automatyki, Wydzia{\l} Automatyki Elektroniki 
   i Informatyki, Politechnika \'{S}laska, Akademicka 16, 44-100 Gliwice, Poland}
\affil[Graz]{Institute of Communication Networks and Satellite Communications, 
   Graz University of Technology, Inffeldgasse 12, 8010 Graz, Austria}
\affil[Vienna]{Institut f\"{u}r Astrophysik, Universit\"{a}t Wien, 
T\"{u}rkenschanzstrasse 17, 1180 Wien, Austria}
\affil[Zagreb]{Department of Physics, Faculty of Science, University of Zagreb, 
Bijeni\v{c}ka cesta 32, 10000 Zagreb, Croatia}

\title{$\beta$ Lyrae as seen by BRITE in 2016
}

\begin{document}

\maketitle

\begin{abstract}
The BTr and UBr satellites observed $\beta$ Lyrae from May to October 2016
to continuously monitor light-curve instabilities with the time resolution 
of  about 100 mins. An instrumental problem affecting localized patches 
on the BTr CCD detector has been discovered by comparison with 
partly simultaneous UBr observations; the origin of the
problem is being investigated. A zero-point offset permits utilization of
the BTr data for a time-series characterization of deviations from the 
mean light curve defined to $\simeq 0.0025$ mag. 
\end{abstract}

\section{Introduction}
\label{intro}

$\beta$ Lyrae (HD~174638) is a bright ($V_{\rm max}=3.4$, $B-V=0.0$) 
eclipsing binary with the orbital period of $P = 12.915$ days,
consisting of a B6-8~II star and an invisible component
which eclipses the B-type star, but otherwise is very hard 
to detect and characterize.
The B-type star has the mass of about $3 M_\odot$ while the estimated
mass of the invisible star is about $13 M_\odot$.
The invisible component is apparently completely shrouded by a 
toroidal accretion disk formed by matter continuously 
transferring from the B-type star. 
The mass transfer results in the period lengthening at an almost
constant rate of 19 seconds per year; as the result, 
the eclipses have shifted in time -- relative to a
linear prediction -- by several orbital cycles during the 
two and half centuries of the photometric observations of $\beta$ Lyrae. 
The complex structure of the system and its extensive studies
are described in a highly readable review by \citet{Harm2002}. 
Among the more recent research of particular note are the
efforts at the interferometric imaging of the binary by the CHARA array by
\citet{Zhao2008} and \citet{Bonn2011} while the latest among several 
$\beta$ Lyrae models are by \citet{Menn2013}. 

Light curves of $\beta$ Lyrae have been known to show a relatively large
scatter. An  international campaign organized in 1959 
\citep{LL1969a,LL1969b} gave indications that the scatter
may reach 0.1 mag (see Fig.~8 in \citet{LL1969a})  
and confirmed that it is not due to 
difficulties with photometric observations of bright stars or to large
angular distances to comparison stars, but is definitely caused by the 
star itself. Later multi-spectral observations and several efforts at
modelling of the binary led to a firm conclusion that the mass transfer 
and accretion processes are unstable and cause photometric 
instabilities. However, those instabilities remain poorly
characterized, mostly because of the moderately long
orbital period and the diurnal interruptions of ground-based observations.
A possibility of a continuous photometric monitoring  over several months
was the driver for the BRITE Constellation observations of 
$\beta$ Lyrae reported here. 

\section{Observations}
\label{obs}

The BRITE Constellation observations of $\beta$ Lyrae were made between 
4 May 2016 and 3 October 2016, for 152 days. Two red-filter satellites,
UniBRITE (UBr) and BRITE-Toronto (BTr) and one blue-filter satellite, 
BRITE-Lem (BLb), were dedicated to these observations. 
Technical difficulties with BLb permitted the blue-filter observations 
for only a fraction of one binary cycle between $t=1635$ and 1647
($t =$ HJD$- 2\,456\,000$)
making these observations useless for the current purpose. 

\begin{figure}
\includegraphics[width=\textwidth]{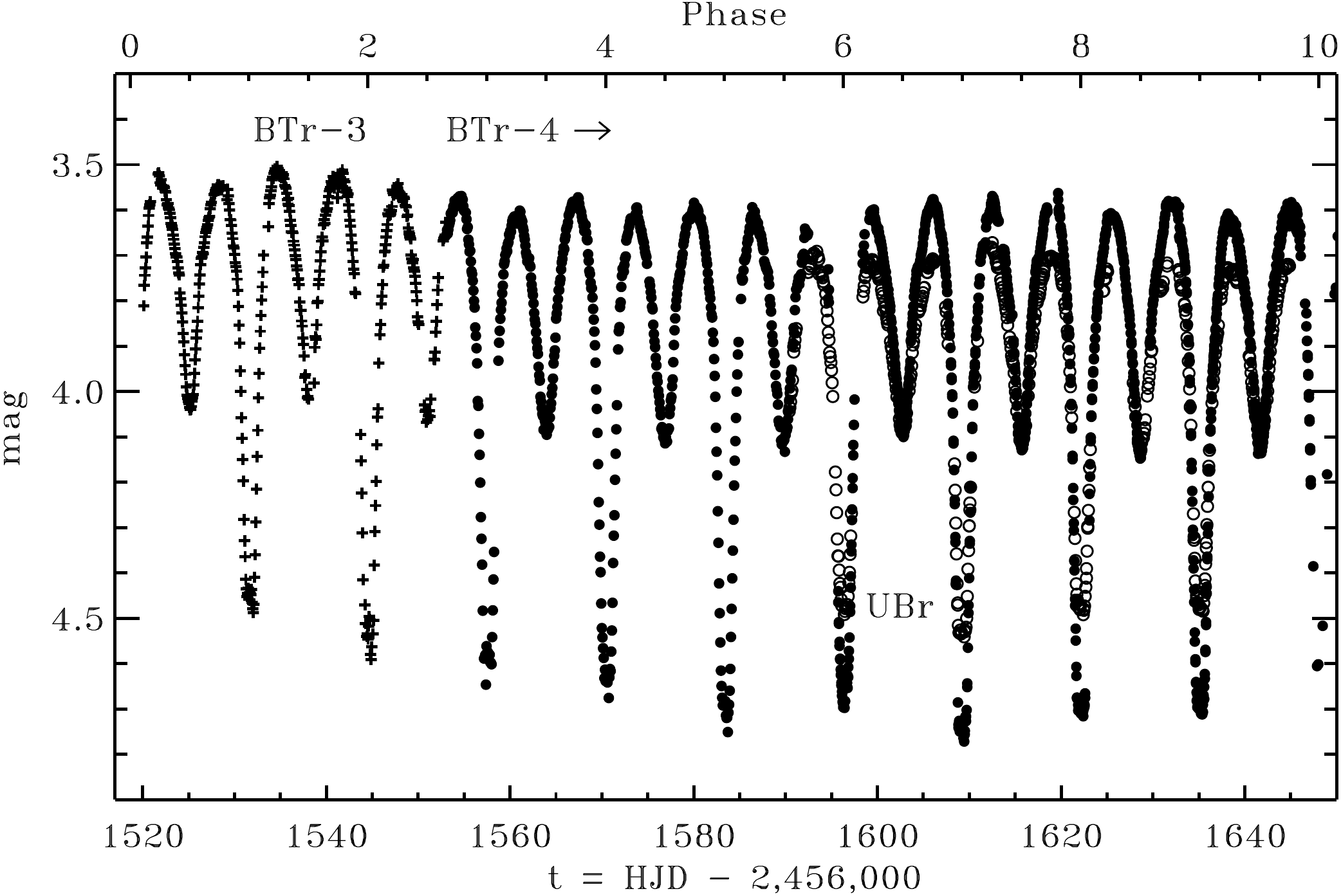}  
\caption{BRITE observations of $\beta$ Lyrae in 2016. The symbols and labels
relate to different setups of the red-filter satellites, as described in the text. 
Note the larger amplitude of the BTr variability compared with that for UBr. 
This difference alerted us about a previously unrecognized CCD 
charge-transfer problem affecting BTr. It may -- locally -- affect the two
remaining red-filter BRITE satellites (Sect.~\ref{prob}).}
\label{fig1_BetaLyr}
\end{figure}

The results for the red-filter satellites are shown in 
Figure~\ref{fig1_BetaLyr} using different symbols  for the satellite-orientation
and the initial image-processing ``setups''. 
Since the setups involved possible zero-point magnitude shifts, 
a decision was made to use only the longest time series designated as BTr-3, BTr-4
and UBr (i.e.\ the combined setups UBr-1 and UBr-2). The BTr observations were 
made with the same exposure time of 3 secs resulting in a median error per a single 
observation of 0.014 mag. For the UBr observations with 1 sec exposures, 
the median error was 0.019 mag.
Most of our analysis has been done with the  orbit averages. 
The satellite-orbit sampling corresponded to 98.27 minutes (BTr) or 100.4 minutes
(UBr).

\section{An instrumental problem}
\label{prob}

The BTr observations provided an excellent record of the variability of 
$\beta$ Lyr sampled almost continuously for 132 days (over 10 binary orbits)
 at 98 minute intervals.
The BTr-3 series extended from $t=1520$ to 1553 while the BTr-4 from 1553 to 
1652. The BTr-4 series overlapped for 50.1 days (3.87 orbital
cycles of the binary) with the UBr observations providing a crucial check of 
consistency of results for both satellites. The check revealed 
a problem:  While the UBr
observations showed an expected amplitude of about 0.85 mag, the
amplitude from the otherwise excellent BTr series
turned out to be larger than one magnitude.
Such large amplitude was never observed before. We had 
no previous indications of any nonlinearities in the BTr data. 
It became obvious that we were seeing a new instrumental 
problem.  

\begin{figure}[h]
\centering
\includegraphics[width=0.40\textwidth]{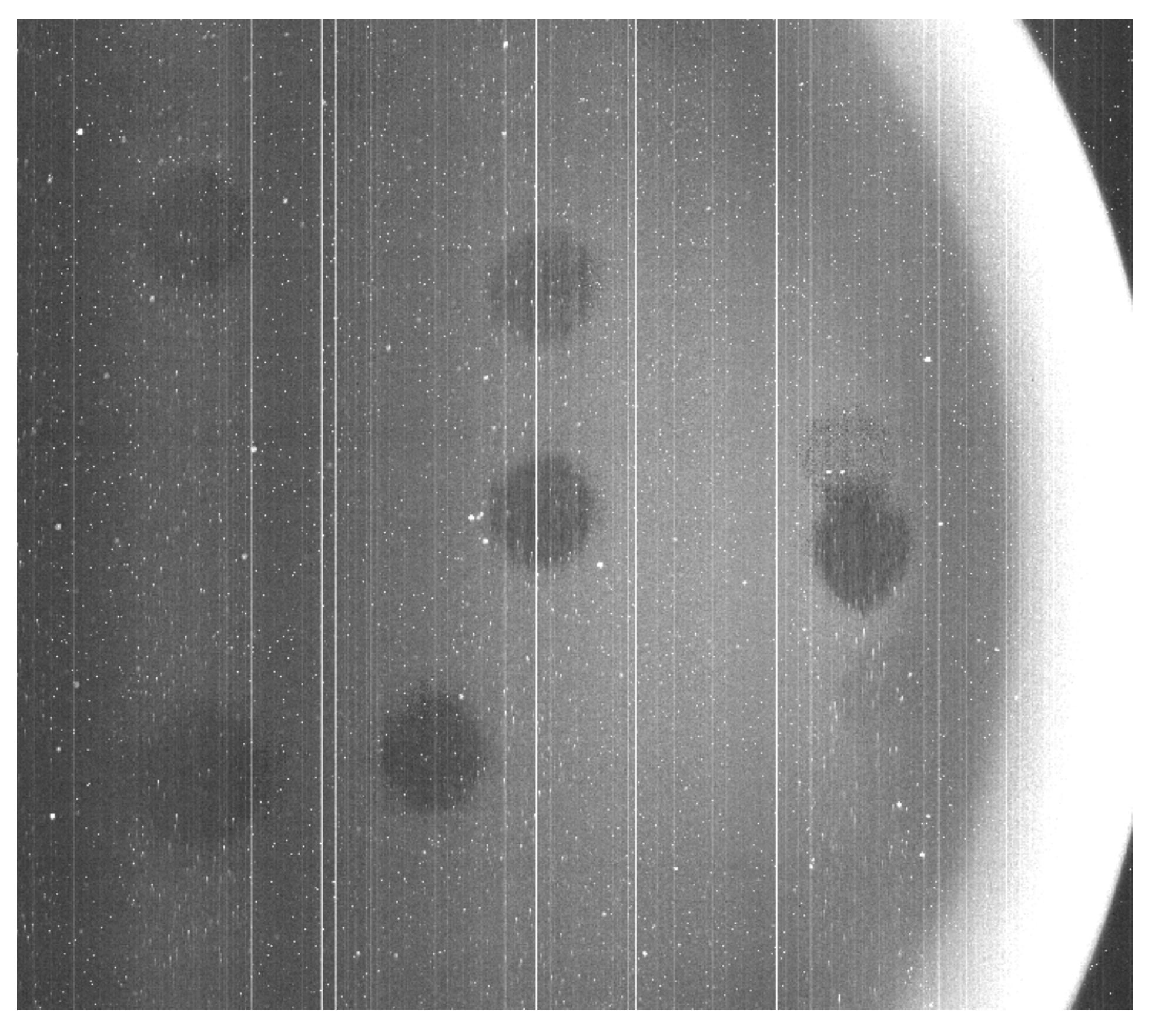}   
\caption{A  section of the BTr full-frame image 
dominated by the Earth-shine background signal. Note the
presence of elliptical signal depressions. They are very shallow and 
thus affect images of faint stars to a stronger degree.  
Similar blemishes appear to be present locally on the CCD detectors 
of the UBr and BHr satellites. 
In this picture, the depressions have been 
made better visible by enhancing the contrast through histogram 
stretching.}
\label{fig2_BetaLyr}
\end{figure}

An investigation of an older, full-frame image from BTr showed that  
the CCD background does show very shallow depressions of similar
shape and size (Fig.~\ref{fig2_BetaLyr}). Apparently, the image of 
$\beta$ Lyr happened to be located in one of such blemishes. 
Further work attempting to characterize the problem has revealed that: 
(1)~All three red-filter satellites show the blemishes to some degree,
but they do not seem to be present in the two blue-filter satellites.
(2)~The blemishes are very similar in terms of their size and shape 
in all three red-filter satellites.
The problem is currently intensely investigated in order to
characterize it and invent methods of its mitigation. 
New, full-frame data are currently being collected for all the satellites.
This is a tedious and slow process and has a detrimental 
impact on the ongoing science operations. All indications are
that it is related to the particle-radiation damage to the CCD
detectors causing charge transfer and trapping effects. However,
it is not clear why only the red-filter satellites are affected and why
the signal-loss areas are so similar in size ($\simeq 200$ pix) and
depth (about 50 ADU).  
It is possible that a cloud of secondary particles is released from 
an optical element (a lens or filter) when it is hit by an energetic, 
primary particle (proton);
the distance between CCD and the optical element would then 
determine the size of the affected region.

An attempt has been made to salvage the otherwise excellent BTr 
data for $\beta$ Lyr. The presence of background blemishes suggested
that some sort of a process modified the signal which -- by itself --
retained linear properties. 
We assumed that the CCD signal was modified by removal  
of a constant amount $C$. This led to a smaller detected signal,
$s(t) = S(t)-C$ with an increased degree of modulation and
thus a larger observed amplitude of $\beta$ Lyr.
This way we could restore $S(t)$ by assuming 
that the amplitude of light variations observed by BTr must be the
same for the overlapping UBr observations; obviously, a second 
assumption was made that observations from UBr had not been affected
by a similar problem. The mapping turned out to be 
excellent with a perfect match of the two fluxes variations.
Thanks to the very large number of the observations we could determine 
the lost signal to be $0.2559 \pm 0.0007$ 
 of the maximum-light $\beta$ Lyr signal.
While the correction applies strictly to the BTr-4 setup, we used it also
for the BTr-3 series because the star remained in the same location 
on the CCD detector. Still, for a study of the light-curve deviations,
the two time series must be done separately
because of the possible small zero-point magnitude shift between the
setups, although a combined, phase-binned light curve (see below) 
indicates a perfect agreement between the two datasets.

\begin{figure}[h!]
\centering
\includegraphics[width=0.62\textwidth]{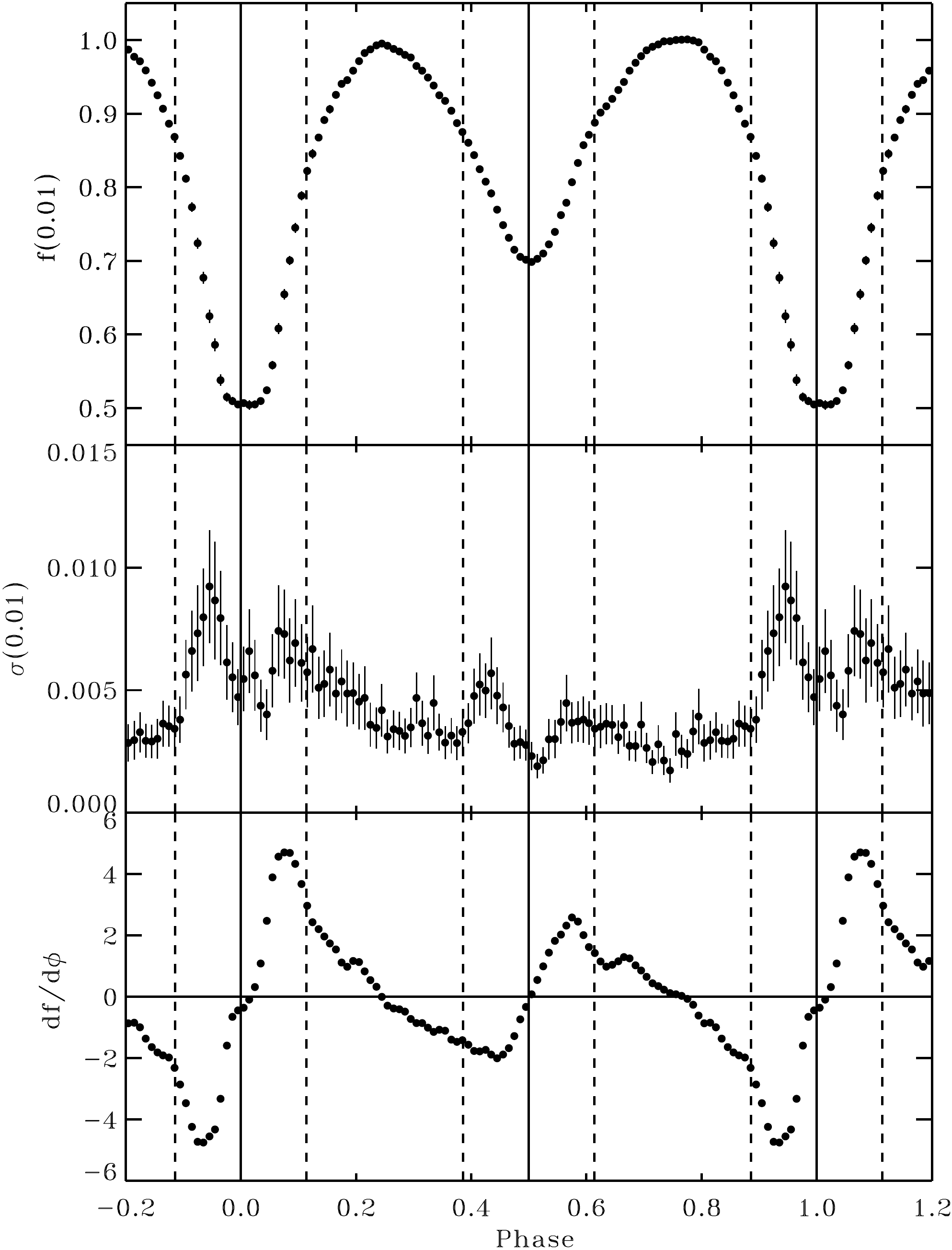}   
\caption{The phase-binned light curve of $\beta$ Lyr formed from the combined
BTr-3 and BTr-4 datasets is expressed here in relative flux units per a 0.01
phase interval (the upper panel). Since epoch-to-epoch averaging is involved,
the errors (the middle panel) are the largest in the eclipse branches,
i.e.\ where the absolute value of the flux derivative
$|df/d\phi|$ is the largest (the lowest panel).
The broken lines indicate the phases of external eclipse
contacts as in the unpublished star-disk model by KP. 
}
\label{fig3_BetaLyr}
\end{figure}

\section{The phase-binned light curve}
\label{lc}

The satellite-orbit averaged data consist of 410 BTr-3, 1230 Btr-4 and
479 UBr observations with the scatter estimated from individual data points
of 0.0017, 0.0013 and 0.0019 mag, respectively (median errors); the small
errors are obviously due to the large number of averaged data points. 
Although deviations from the mean, phase-binned BTr light 
curve (Fig.~\ref{fig3_BetaLyr}) reach $\pm 0.04$ mag, the median errors
per 0.01 phase interval are typically at the level of 0.002 -- 0.003 mag. 
Such a well defined light curve appears to be of interest for 
improvement of the $\beta$ Lyr system models; a preliminary use of 
the current light curve was presented by KP at the 
EWASS Conference\footnote{European Week of Astronomy and Space Science,
Prague, 26-30 June 2017 (no publication).}.
The flux derivative $df/d\phi$ computed using the light curve
is smooth so that its inflection points confirm the predicted 
eclipse contact phases. It is interesting that small, but fairly well defined 
additional notches in the derivative seem 
to appear slightly beyond the contact phases, at approximately 
similar distances; they may indicate the presence of matter distributed 
in the system above the disk/star surfaces.

\section{Conclusions}
\label{concl}

The instrumental problem that we detected in the BTr data in the 
$\beta$ Lyr observations
is currently investigated as it may be occurring in all three red-filter satellites.
A simple, linear signal transformation has permitted to correct the time
series of BTr data. This time series is unique as it extends 
for over 10 orbital cycles of a continuous and frequent ($\simeq 98$ min)
time sampling. 
The data will be the subject of a time-series analysis while the mean
light curve will be used for a $\beta$ Lyr modelling.

\acknowledgements{Thanks are due 
to Dr.\ Dietrich Baade for reminding the first author about the 
well-known effect pertaining to the noise in averaged light curves.

SMR acknowledges the financial support from NSERC (Canada). 
The support  from the Polish National Science Centre (NCN) 
is acknowledged by APi (grant 2016/21/B/ST9/01126) and
APo (grant 2016/21/D/ ST9/00656).
KP has been supported in part by the Croatian Science
Foundation under the grant 2014-09-8656. 
}

\bibliographystyle{ptapap}
\bibliography{BetaLyr_Slavek}

\end{document}